\documentclass[twocolumn,twoside,slac_two]{revtex4}
\usepackage{graphicx}
\usepackage{fancyhdr}
\begin{document}

\title{All-Sky Earth Occultation Observations with the Fermi Gamma Ray Burst Monitor}

%

\author{Colleen A. Wilson-Hodge}
\affiliation{NASA Marshall Space Flight Center, Huntsville, AL 35812, USA}
\author{E Beklen}
\affiliation{Middle East Technical University, 06531, Ankara, Turkey}
\author{P.N. Bhat, M.S. Briggs, V. Chaplin, V. Connaughton}
\affiliation{University of Alabama in Huntsville, Huntsville, AL 35899, USA}
\author{A. Camero-Arranz}
\affiliation{Fundaci\'{o}n Espa\~{n}ola de Ciencia y Tecnolog\'{i}a (MICINN), C/Rosario Pino 14-16, 28020, Madrid, Spain}
\author{G. Case, M.Cherry, J. Rodi}
\affiliation{Lousiana State University, Baton Rouge, LA, 70803 USA}
\author{M.H. Finger}
\affiliation{Universities Space Research Association, Huntsville, AL 35806, USA}
\author{P. Jenke}
\affiliation{NASA Postdoctoral Program Fellow, NASA Marshall Space Flight Center, Huntsville, AL 35812,
USA}
\author{R.H. Haynes}
\affiliation{NASA Academy Program Fellow, NASA Marshall Space Flight Center, Huntsville, AL 35812, USA}

\begin{abstract}
Using the Gamma Ray Burst Monitor (GBM) on-board Fermi, we are monitoring the hard X-ray/soft gamma ray sky using the Earth occultation technique. Each time a source in our catalog  enters or exits occultation by the Earth, we measure its flux using the change in count rates due to the occultation. Currently we are using CTIME data with 8 energy channels spanning 8 keV to 1 MeV for the GBM NaI detectors and spanning 150 keV to 40 MeV for the GBM BGO detectors. Our preliminary catalog consists of galactic X-ray binaries, the Crab Nebula, and active galactic nuclei. In addition, to Earth occultations, we have observed numerous occultations with Fermi's solar panels. We will present early results. Regularly updated results can be found on our website \url{http://gammaray.nsstc.nasa.gov/gbm/science/occultation}.

\end{abstract}

\maketitle

\thispagestyle{fancy}


\section{INTRODUCTION}

\subsection{ Fermi Gamma Ray Burst Monitor}

The Gamma Ray Burst Monitor (GBM) on board {\it Fermi} consists of twelve Sodium
Iodide (NaI) detectors and two Bismuth Germinate (BGO) detectors, sensitive from
about 8 keV to 1 MeV and 150 keV to 40 MeV, respectively. Continuous data from both detector types are
received from GBM when the spacecraft is outside the South Atlantic Anomaly.
They consist of two types, CTIME, with 256 ms nominal time resolution and eight
energy channels, and CSPEC, with 4.096 s nominal time resolution and 128 energy
channels \cite{Meegan09}. For the NaI detectors the lower energy bounds of the 8 CTIME bands are: 8, 12, 25, 50, 100, 300,
500, and 1000 keV. To date the Earth occultation technique has used only CTIME data, but
future plans include applying the technique to CSPEC data as well.

\subsection{Scientific Objectives}

The goals of the GBM Earth Occultation Project are to:
\begin{itemize}
\setlength{\baselineskip}{10pt}
\setlength{\parskip}{-2pt}
\item Search for rapid bright events such as AGN flares, rapid transients, black hole transient outburst onsets, and bright flares in known transient objects,
\item Provide light curves and spectra of these events to the scientific community and to the {\it Fermi} Large Area Telescope (LAT) team to allow correlations with LAT observations and the results of correlated  multiwavelength observations,
\item Monitor the majority of the soft gamma ray sky including known transient and variable objects such as galactic black hole transients, neutron star binaries, and AGNs, looking for long term changes and bright outbursts,
\item Discover and study new transient objects and provide alerts to other instruments for more sensitive follow-up searches, and
\item Reveal pre-discovery emission from sources that are newly discovered by {\it Fermi} and/or other observers.
\end{itemize}

\section{EARTH OCCULTATION TECHNIQUE}

The Earth occultation technique is currently used with GBM to monitor fluxes
for a catalog of known sources. Sources are added to the catalog as they flare
or upon request. For each source, the time of 50\% atmospheric transmission at
100 keV is predicted using the known position of {\it Fermi}. A four-minute
window of data is selected centered on the predicted time. NaI detectors
are selected if the angle between the detector normal and the source is
less than $60^\circ$. The data in each energy channel for each detector are then fitted with a model consisting of 
a quadratic background and a scale factor times a model profile for each source. Model profiles for the source of 
interest and any other sources occulting within the fit window are computed by convolving the atmospheric 
transmission and detector response for each data point with an assumed energy spectrum for the source. The profile 
for each source is normalized using the count rate predicted by convolving the detector response at the occultation 
time with the assumed source spectrum (i.e., assuming no atmospheric transmission). Once scale factors are fitted for
all detectors, a fit across detectors is then performed to find the best scale factor for all detectors. This scale
factor is multiplied by the assumed flux model integrated across each energy channel to determine flux for each
source.

This method differs considerably from the Earth occultation technique used with the Burst and Transient Source
Experiment (BATSE) \cite{Harmon02, Harmon04} on the {\it Compton Gamma Ray Observatory (CGRO)}. {\it Fermi} is pointed
$\pm (35-50)^{\circ}\ $ from zenith, so the detector response relative to a particular source is constantly changing
and must be accounted for within each individual step fit. {\it CGRO} remained at a fixed pointing relative to the
sky for typically two weeks, so the detector responses could be ignored for the individual step fits and needed to be
applied only to compute fluxes. GBM has additional energy coverage in the 8-25 keV band, at lower energies than BATSE
could detect. The placement of the GBM detectors on the spacecraft also results in occasional blockages by the solar
panels and other parts of the spacecraft. These must be accounted for in the detector response as well. Currently the
solar panels are not well modeled in the detector response because they are constantly moving. If a detector is
blocked by a solar panel during a fit window, it is excluded from the fit.

\section{SOURCE MONITORING RESULTS}

As of December 15, 2009, we are currently monitoring a catalog of 64 sources (60 X-ray binaries, 3 AGN, and the
Crab) with GBM Earth occultation. For 49 of these sources Earth occultation measurements 
have been made from August 12, 2008 until December 15, 2009, and preliminary analysis has been performed. Preliminary 
detections are listed in Table~1: Y denotes $>10 \sigma$ statistical significance for the 
mission-long average in at least one energy channel (37 sources), T denotes transient activity consistent with 
Swift/BAT (5 sources). Sources added recently, with incomplete histories and analysis are listed in Table~2. Sources not 
currently marked as detections, especially transients, may be detectable with further analysis. Consistency checks were performed by 
comparing GBM and Swift/BAT Hard X-ray Transient Monitor 
\footnote{\url{http://swift.gsfc.nasa.gov/docs/swift/results/transients}} measurements normalized
the Crab. Light curves are shown for selected sources below. Systematic problems due to interfering sources are still present at times in 
the light curves. Optimizing the list of sources to be considered in step fitting is an iterative process and these effects will likely 
improve with future runs through the GBM data.

\subsection{Recent Transients}

\begin{figure}
\hspace{-15mm}\includegraphics[width=85mm]{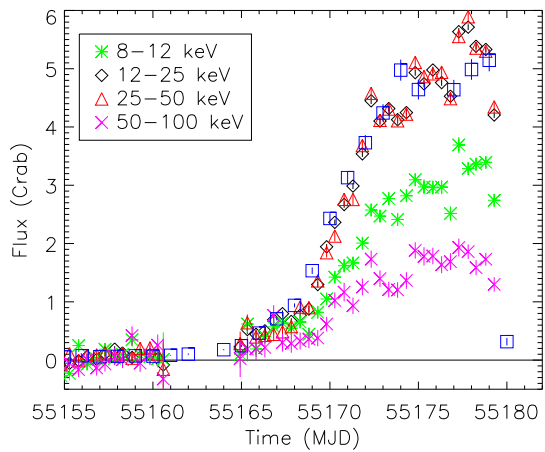}
\caption{A0535+262 giant outburst. Green asterisks, black diamonds, red triangles, and pink crosses denote 8-12, 12-25, 25-50, 50-100 keV GBM fluxes
averaged over 1/2 day, respectively. Blue squares denote 1-day Swift/BAT 15-50keV fluxes.}
\label{a0535_giant}
\end{figure}

\begin{figure}
\hspace{-15mm}\includegraphics[width=85mm]{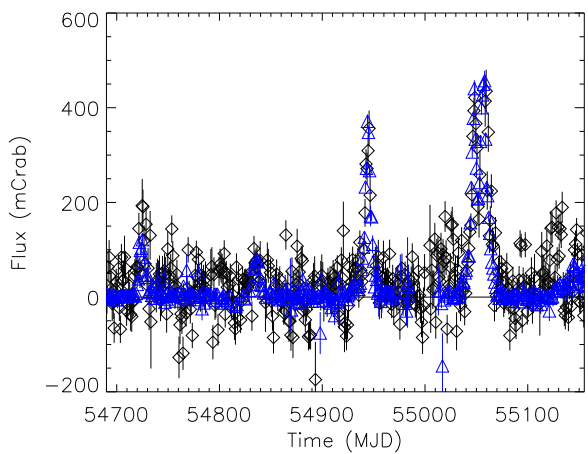}
\caption{A0535+262 normal outbursts observed with GBM. Black diamonds and blue squares denote GBM 12-50 keV and Swift/BAT 15-50 keV measurements,
respectively.}
\label{a0535_normal}
\end{figure}

{\bf A0535+262}, an 103-second Be/X-ray transient accreting pulsar, began a giant outburst in late November 2009 \cite{WilsonHodge09}. The 
flux rose rapidly to nearly 6 Crab in the 12-25 keV and 25-50 keV bands and about 2-3 Crab in the 8-12 keV and 50-100 keV bands on 
December 10, 2009. The light curve of this outburst is shown in Figure~\ref{a0535_giant}. The 12-25 and 25-50 keV
bands track each other closely and both agree well with Swift/BAT. The A0535+262 giant outburst is not detected in daily averages in the 100-300 keV band. 
Since {\it Fermi} was launched, 
A0535+262 underwent four normal outbursts prior to the giant outburst. Figure~\ref{a0535_normal} shows the comparison between GBM
and Swift/BAT data for these outbursts. Pulsations from A0535+262 are detected with the GBM pulsar monitoring technique \cite{Finger09}.

\begin{figure}
\hspace{-15mm}\includegraphics[width=85mm]{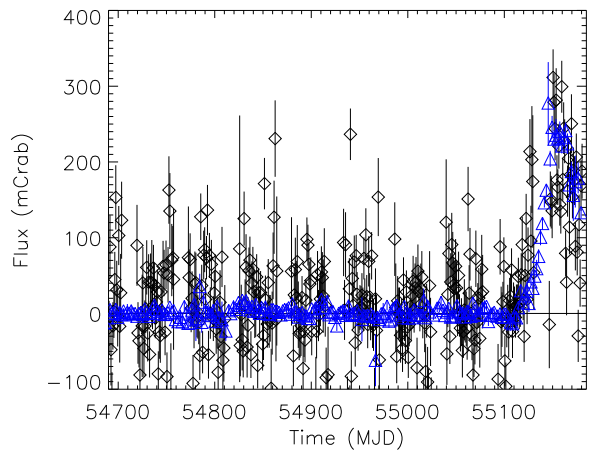}
\caption{H1417--624 long-term flux history. A giant outburst is visible near the end of the light curve. Black diamonds and blue squares denote GBM 12-50 keV and Swift/BAT 15-50 keV measurements,
respectively.} 
\label{h1417}
\end{figure}

{\bf H1417--624} is Be/X-ray transient pulsar with a 17.5 s period
\cite{Beklen09} that began a giant outburst in late October 2009. GBM detects
H1417--624 in the 12-25 and 25-50 keV bands, which are summed and compared with
Swift/BAT in Figure ~\ref{h1417}. Considerable scatter is visible in the H1417--624 light curve prior to the giant outburst. This is likely due to
nearby nsources that undergo Earth occultation at close to the same time as H1417--624. the  Pulsations from H1417--624 are detected by the GBM 
pulsar monitoring technique \cite{Finger09}.

\begin{figure}
\hspace{-15mm}\includegraphics[width=85mm]{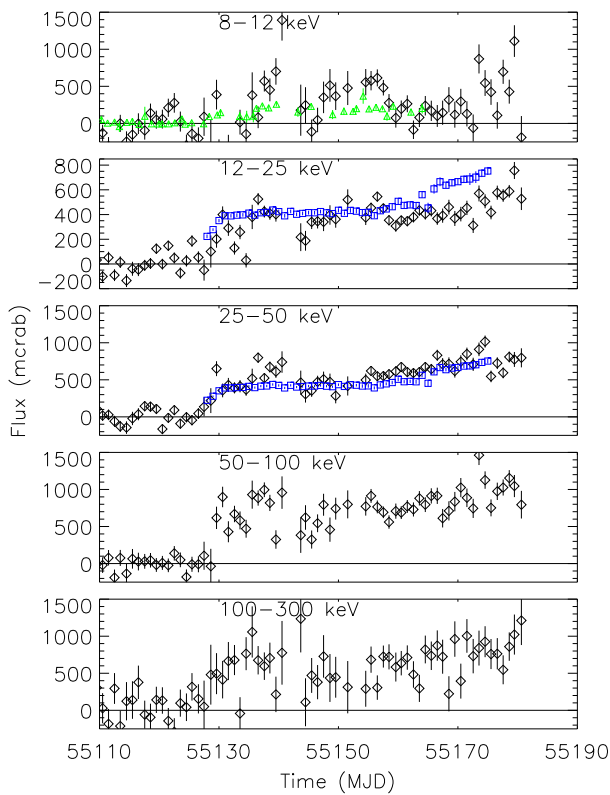}
\caption{XTE J1752--223 outburst observed with GBM in 5 energy bands 8-12, 12-25, 25-50, 50-100, and 100-300 keV. RXTE/ASM 5-12 keV fluxes are
overplotted for comparison in the 8-12 keV band. Swift/BAT 15-50 keV measurements are
overplotted as blue squares for comparison to the GBM 12-25 and 25-50 keV bands.}
\label{xtej1752}
\end{figure}

{\bf XTE J1752-223}, a new black hole transient discovered on October 23, 2009, with the {\it Rossi X-ray Timing Explorer (RXTE)} \cite{Markwardt09}.
Figure~\ref{xtej1752} shows the GBM light curves in 5 energy bands, spanning 8-300 keV. This transient is one of seven sources
detected with GBM above 100 keV \cite{Cherry09}. XTE J1752--223 rose to a plateau in all bands from 12-300 keV
for about 30 days. Then the source began brightening again. From examining the GBM light curves, this increase appears stronger in the 25-50 keV band
than in the 12-25 keV range, suggesting possible spectral changes.

\subsection{Persistent Sources}

\begin{figure}
\hspace{-15mm}\includegraphics[width=85mm]{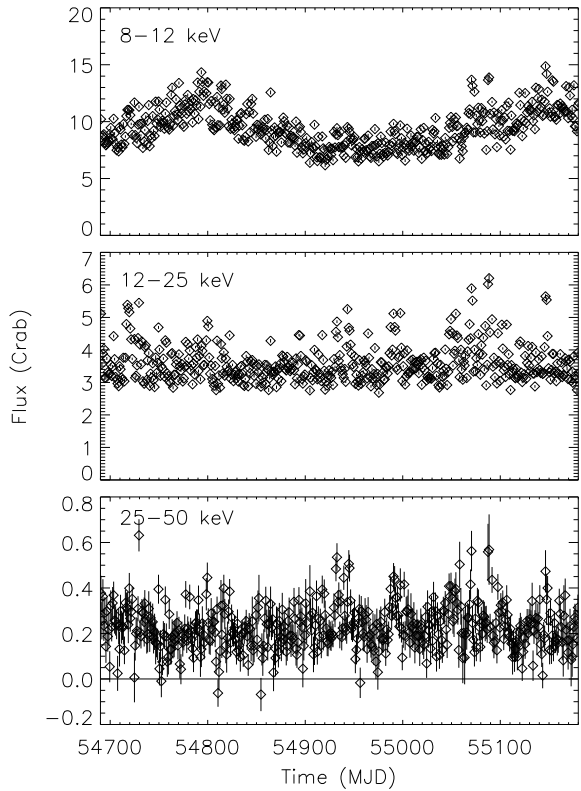}
\caption{Sco X-1 GBM daily flux averages in the 8-12 (top), 12-25 (middle), and 25-50 keV (bottom) ranges.}
\label{scox1}
\end{figure}

{\bf Sco X-1} is a bright Z-source. GBM detects it easily in less than a day in the 8-12, 12-25, and 25-50 keV bands. Sco X-1 is so
bright that we have the potential to use single occultation steps for analysis. Figure~\ref{scox1} shows daily
average fluxes.

\begin{figure}
\hspace{-15mm}\includegraphics[width=85mm]{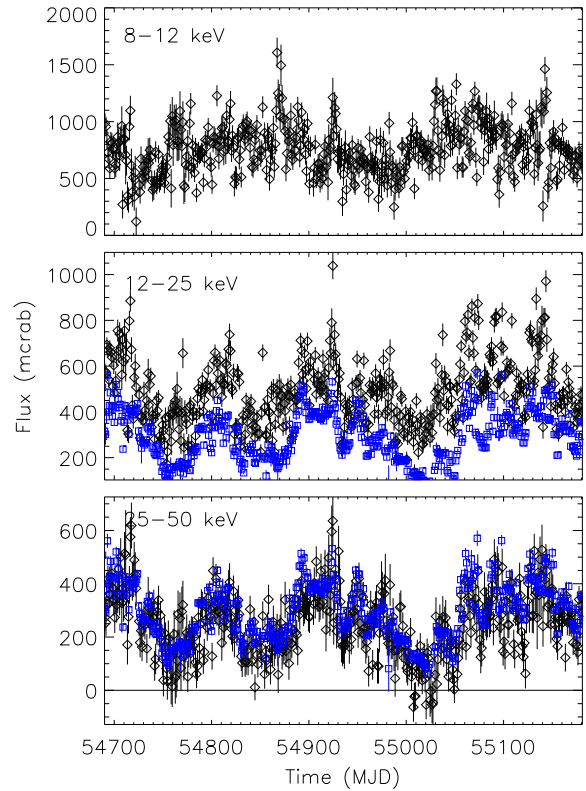}
\caption{GRS 1915+105 GBM light curves in the 8-12 (top), 12-25 (middle), and 25-50 keV (bottom) bands. Higher energy light curves are shown in Cherry et al. 
(2009). Black diamonds denote GBM measurements and blue squares denote Swift/BAT 15-50 keV measurements.}
\label{grs1915}
\end{figure}

{\bf GRS 1915+105} is a microquasar. GBM detects GRS 1915+105 on a daily basis in the 8-12, 12-25, and 25-50 keV, and 50-100 keV and on longer timescales in the
100-300 keV band \cite{Cherry09}. Figure~\ref{grs1915} shows three lower energy bands.

\begin{figure}
\hspace{-15mm}\includegraphics[width=85mm]{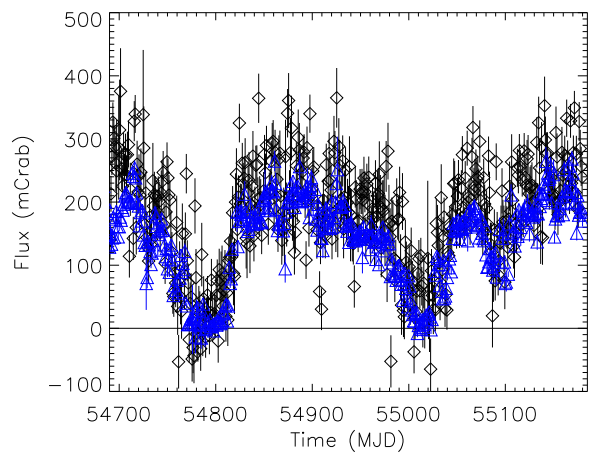}
\caption{CYG X-3 GBM light curve in the 12-50 keV range. Black points are GBM 1-day averages. Blue points are Swift/BAT 15-50 keV
measurements.}
\label{cygx3}
\end{figure}

{\bf Cyg X-3} is a probable black-hole system. GBM detects Cyg X-3 on a daily basis in the 8-12, 12-25, and 25-50 keV bands. State changes are detected with GBM and agree well with Swift/BAT, as shown in Figure~\ref{cygx3}.

\subsection{Extra-galactic Sources}

\begin{figure}
\hspace{-15mm}\includegraphics[width=85mm]{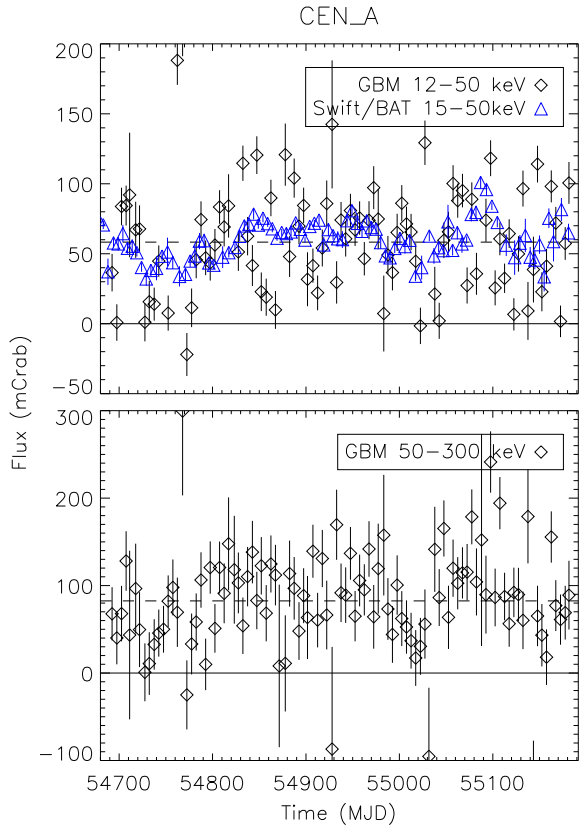}
\caption{Cen A light curves in the 12-50 and 50-300 keV ranges. Black points denote GBM 5-day averages. Blue points denote 5-day average Swift/BAT 15-50 keV measurements.
Dashed lines denote the average flux level in GBM.}
\label{cena}
\end{figure}

Using the Earth occultation technique, we have monitored Cen A since August 2008. Cen
A is detected on a daily basis in the 12-25 keV band and over longer timescales in the 25-50, 50-100, and 100-300 keV bands. See Figure~\ref{cena} and Cherry et al. (2009) for more discussion. Two
additional extra-galactic sources, Mrk 421 and 3C 454.3 have recently been added to the occultation catalog and monitoring is underway. As we iteratively improve our bright
source list of X-ray binaries, we will be able to better investigate fainter sources such as active galaxies and search for flaring activity.

\section{Earth Occultation Monitoring Using the BGO Detectors}

\begin{figure}
\hspace{-15mm}\includegraphics[width=85mm]{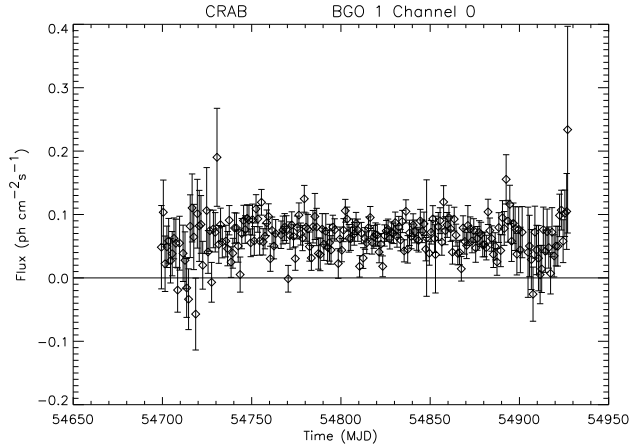}
\caption{Daily average Earth occultation flux measurements for the Crab in BGO 1 in the 100-380 keV band.}
\label{crab_bgo}
\end{figure}

The Earth occultation technique can be applied to the higher energy BGO detectors on GBM, essentially using identical software as for the NaI detectors. The BGO energy
channel edges differ more significantly than the NaI edges, so fluxes are computed for the BGO detectors separately. The lower energy bounds of the CTIME channels are for BGO
0: 0.18, 0.34, 0.97, 2.1, 4.6, 9.8, 21, and 45 MeV and for BGO 1 are: 0.10, 0.38, 0.93, 2.1, 4.8, 10, 22, 46 MeV. In addition, many of the sources that are very
bright that must be considered as interfering sources in the NaIs, e.g. Sco X-1, are not at all bright in the BGOs, so source interference must be addressed carefully
as well. Code to monitor sources using the BGOs has just been developed.  A light curve for the Crab is shown in
Figure~\ref{crab_bgo} for the 100-380 keV band in BGO 1, to demonstrate that preliminary runs have been successful.

\section{SUMMARY}

The GBM Earth occultation technique is working well. Agreement with Swift/BAT is good in the 15-50 keV range.
Currently 64 sources are being monitored. Of these, we have preliminary detections for 42 sources in at least one of
the energy bands between 8 keV and 300 keV. Future plans for the Earth occultation project include monitoring sources using the BGO detectors, expanding the catalog to
include more extra-galactic sources, and application of the technique to CSPEC data.

\bigskip 
\begin{acknowledgments}
The GBM Earth occultation project is a funded Fermi Guest Investigation. This paper uses quicklook results provided by the ASM/RXTE team and Swift/BAT transient monitor results provided by the Swift/BAT team.
\end{acknowledgments}

\bigskip 

\begin{thebibliography}{9}   
\bibitem{Beklen09}
E. Beklen \& M.H. Finger, ATel \#3276 (2009).
\bibitem{Cherry09}
M. Cherry et al., Proceedings of the Fermi Symposium, eConf C091122 (2009).
\bibitem{Harmon02}
B.A. Harmon et al. ApJS, 138, 149 (2002).
\bibitem{Harmon04}
B.A. Harmon et al. ApJS, 154, 585 (2004).
\bibitem{Finger09}
M.H. Finger et al., Proceedings of the Fermi Symposium, eConf C091122 (2009).
\bibitem{Markwardt09}
C.B.Markwardt et al., ATel \#2258 (2009).
\bibitem{Meegan09}
C.A. Meegan et al., ApJ, 702, 791 (2009).
\bibitem{WilsonHodge09}
C.A. Wilson-Hodge, M.H. Finger, A. Camero-Arranz, V. Connaughton, ATel \#2324, (2009).



\end{thebibliography}

\begin{table*}
\hspace{0.033\textwidth}
\begin{minipage}{0.4\textwidth}
\caption{Preliminary Detections as of December 14, 2009}
\footnotesize
\begin{tabular}{|l|l|l|l|}
\hline \textbf{Name} & \textbf{R.A.} & \textbf{Decl.} & \textbf{Preliminary}\\
                     & \textbf{(deg)} & \textbf{(deg)} & \textbf{Detection} \\
\hline V0332+53         &      53.7500&      53.1730 & Y\\
\hline CRAB             &      83.6330&      22.0150 & Y\\
\hline A0535+262        &      84.7280&      26.3160 & Y\\
\hline MXB0656-072      &      104.612&     -7.26300 & T \\
\hline VELAX-1          &      135.529&     -40.5550 & Y \\
\hline MRK421           &      166.114&      38.2090 & Y\\
\hline GROJ1008-57      &      152.442&     -58.2930 & T \\
\hline A1118-61         &      170.238&     -61.9170 & T\\
\hline CENX-3           &      170.314&     -60.6230 & Y\\
\hline 1E1145.1-6141    &      176.869&     -61.9537 & \\
\hline GX301-2          &      186.657&     -62.7700 & Y\\
\hline GX304-1          &      195.321&     -61.6018 & \\
\hline CEN A            &      201.365&     -43.0190 & Y\\
\hline H1417-624        &      215.300&     -62.7000 & Y\\
\hline CENX-4           &      224.592&     -31.6690 & \\
\hline SWIFTJ1539.2-6227&      234.818&     -62.4590 & T\\
\hline H1608-522        &      243.175&     -52.4170 & \\
\hline SCOX-1           &      244.980&     -15.6400 & Y\\
\hline IGRJ16318-4848   &      247.967&     -48.8030 & Y\\
\hline AXJ1631.9-4752   &      248.008&     -47.8740 & Y\\
\hline 4U1636-536       &      250.231&     -53.7510 & Y \\
\hline GX340+0         &      251.449&     -45.6111 & Y \\
\hline HERX-1           &      254.457&      35.3420 & Y\\
\hline OAO1657-415      &      255.199&     -41.6730 & Y\\
\hline GX339-4          &      255.706&     -48.7890 & \\
\hline 4U1700-377       &      255.986&     -37.8440 & Y\\
\hline GX349+2          &      256.450&     -36.4170 & Y\\
\hline 4U1702-429       &      256.564&     -43.0361 & Y\\
\hline H1705-440       &      257.228&     -44.1010 & Y\\
\hline GX354-0          &      263.000&     -33.8330 & Y\\
\hline GX1+4            &      263.008&     -24.7470 & Y\\
\hline 1E1740-29        &      265.969&     -29.7420 & Y \\
\hline 1A1742-294       &      266.523&     -29.5150 & Y\\
\hline IGRJ17464-3213   &      266.575&     -32.2400 & Y\\
\hline IGRJ17473-2721   &      266.825&     -27.3440 & Y\\
\hline XTEJ1752-223     &      268.044&     -22.3250 & Y\\
\hline SWIFTJ1753.5-0127&      268.368&     -1.45300 & Y\\
\hline GX5-1            &      270.275&     -25.0830 & Y\\
\hline GX17+2           &      274.006&     -14.0360 & Y\\
\hline H1820-303       &      275.925&     -30.3670 & Y\\
\hline SWIFTJ1842.5-1124&      280.500&     -11.4000 &  \\
\hline GS1843+00        &      281.412&     0.891000 & Y\\
\hline AQLX-1           &      287.817&     0.584900 & \\
\hline GRS1915+105      &      288.820&      10.9700 & Y\\
\hline CYGX-1           &      299.590&      35.2020 & Y\\
\hline EXO2030+375      &      308.064&      37.6390 & Y\\
\hline CYGX-3           &      308.107&      40.9580 & Y\\
\hline GINGA2138+56     &      324.878&      56.9861 & T\\
\hline CYGX-2           &      326.170&      38.3200 & Y\\
\hline
\end{tabular}
\label{catalog}
\end{minipage}
\hspace{0.066\textwidth}
\begin{minipage}{0.4\textwidth}
\vspace{-5.75in}
\centering
\caption{Recently Added Sources}
\footnotesize
\begin{tabular}{|l|l|l|}
\hline \textbf{Name} & \textbf{R.A.} & \textbf{Decl.} \\
                     & \textbf{(deg)} & \textbf{(deg)} \\
\hline LSI+61 303       &      40.1319&      61.2293 \\
\hline 4U1538-52       &      235.597&     -52.3860 \\
\hline X1624-490       &      247.011&     -49.1983  \\
\hline 4U1626-67       &      248.070&     -67.4620  \\
\hline SWIFTJ1713.4-4219&      258.361&     -42.3270 \\
\hline GX9+9           &      262.934&     -16.9614 \\
\hline X1735-444        &      264.743&     -44.4500  \\
\hline GX3+1           &      266.984&     -26.5636 \\
\hline GX9+1           &      270.382&     -20.5310 \\
\hline GX13+1         &      273.632&     -17.1559 \\
\hline GS1826-238      &      277.367&     -23.7970 \\
\hline SERX-1          &      279.990&      5.03560 \\
\hline HT1900.1-2455    &      285.036&     -24.9205 \\
\hline 4U1954+31        &      298.929&      32.1000 \\
\hline 3C454.3         &      343.490&      16.1480 \\
\hline
\end{tabular}
\label{recent}
\end{minipage}
\end{table*}


\end{document}